# Enhanced strain rate sensitivity due to platelet linear complexions in Al-Cu

Pulkit Garg [a,b], Daniel S. Gianola [c], Timothy J. Rupert [a,d,e,*]
[a] Department of Materials Science and Engineering, University of California, Irvine, CA 92697, USA
[b] Department of Mechanical Engineering, University of California, Santa Barbara, 93106-5070, CA, USA
[c] Materials Department, University of California, Santa Barbara, CA 93106, USA
[d] Hopkins Extreme Materials Institute, Johns Hopkins University, Baltimore, MD 21218, USA
[e] Department of Materials Science and Engineering, Johns Hopkins University, Baltimore, MD 21218, USA

* To whom correspondence should be addressed: tim.rupert@jhu.edu

**Abstract**

Platelet array linear complexions have been predicted in Al-Cu, with notable features being dislocation faceting and climb into the precipitate, both of which should impact plasticity. In this study, we examine the strain rate dependence of strength for platelet linear complexions using atomistic simulations, with classical precipitate strengthening through particle cutting and particle bowing used as baseline comparisons. Dislocation segments with edge character must climb down from the platelet structures prior to the commencement of glide, introducing a significant time-dependent barrier to plastic deformation. Consequently, the strain rate sensitivity of strength for the platelet linear complexions system was found to be up to five times higher than that of classical precipitation strengthening mechanisms.

**Keywords:** Dislocations; Strain rate sensitivity; Precipitation; Aluminum alloys, Molecular dynamics (MD)

Linear complexions (LCs) are recently discovered defect states where the structure and chemistry near a dislocation varies due to the local distortion field [1]. These features represent a new pathway to manipulate microstructure and modify the mechanical behavior of engineering alloys, as the defects that control plasticity are directly modified. LCs were first studied in body-centered cubic alloys such as Fe-Mn [2, 3], Fe-Ni [4-6], and reactor pressure vessel steels [7], yet more recent work has identified a range of possible LC types in face-centered cubic (FCC) alloys [8, 9]. Among these, Al-Cu alloys are predicted to host platelet-shaped precipitates that grow from the dislocations, moving out of the original slip plane [8] and resembling classical Guinier-Preston (GP) zones found during aging of bulk Al-Cu alloys [8, 10, 11]. Garg and Rupert [12] recently showed that such platelet array LCs restrict dislocation motion and significantly increase strength, yet follow strength-scaling laws that are different from classical precipitate interaction mechanisms. Interestingly, the platelet array LCs were observed to form a complex configuration, where the dislocation line became faceted and edge character segments climbed into the nanoscale precipitates.

In addition to affecting strength, the introduction of new deformation mechanisms can impact the strain rate sensitivity (SRS) of a material. Generally, higher yield and flow stresses are observed as applied strain rate is increased, as less time is given for thermally-activated deformation mechanisms to operate and the system must therefore build higher stresses to drive plasticity. FCC metals typically have low SRS parameter (*m*) values at ambient conditions, as dislocation glide is easily accomplished, Pierels stresses are low, and long-range interactions dominate [13, 14]. For example, Khan and Liu [15] studied the deformation behavior of Al-Cu-Mn alloys at a variety of strain rates ranging from $10^{-4}$ to $10^{3}$ $s^{-1}$ and observed negligible SRS at room temperature, with *m* values typically reported in the range of 0.005-0.02 [16]. A negative

SRS has even been observed for Al-Mg alloys under quasi-static conditions, commonly a result of dynamic strain aging [17-19]. Hue et al. [20] explored the effect of varying strain rate on the strength of Al alloys with atomistic simulations and measured *m* values of 0.037 and 0.044 for strain rates below and above $10^{10}$ $s^{-1}$, respectively . Recent work from Fan et al. [21] demonstrated with discrete dislocation dynamics and molecular dynamics (MD) simulations that SRS can be altered through modification of the dislocation density, showing a deep connection between deformation physics, strain rate, and dislocation density. The SRS of Al-rich alloys has been observed to increase at elevated temperatures [15], with dislocation climb associated with power law creep being one of the possible mechanistic explanations [22].

The introduction of dislocation climb as a rate-limiting plasticity mechanism in Al-Cu with platelet array LCs offers a potential pathway for achieving increased room temperature SRS. Here, such a possibility is investigated using MD simulations of the initial dislocation breakaway event from platelet LC configurations. The SRS of Al-Cu reinforced with platelet LCs is found to be 4-5.5 times higher than the SRS of classical precipitation strengthening mechanisms such as particle bowing or cutting. The dislocation develops facets in order to form edge character segments, which climb into the platelet structure and therefore require climb in the opposite direction before plastic deformation can commence. Overall, this work demonstrates that platelet LCs induce unique deformation mechanisms that can increase the resistance to high strain rate deformation, opening the door for future alloys with extreme strength under dynamic loading and prolonged extents of uniform elongation even under quasi-static conditions.

Models to simulate LC-type and classical precipitate interactions in Al-Cu were created and atomistic simulations were performed using the Large-scale Atomic/Molecular Massively Parallel Simulator (LAMMPS) code [23] with a 1 fs integration time step. All atomic configurations were analyzed and visualized using the common neighbor analysis (CNA) [24] and dislocation analysis (DXA) [25] methods within the visualization tool OVITO [26]. The FCC Al atoms have been removed for clarity in all figures here. First, a platelet LC sample was created using hybrid Monte Carlo (MC)/MD simulations with an embedded atom method potential for the Al-Cu system [27], following the procedures of Ref. [8]. Briefly, a pair of edge dislocations was relaxed and Cu segregation occurred to the compression side of the defects during the MC/MD procedure until platelet-shaped particles were formed. The equilibrated LC configuration at 250 K was obtained using a Nose–Hoover thermos/barostat at zero pressure, yielding stable control with fluctuations within ±20 K in temperature and ±5 MPa in pressure. This interatomic potential was developed to reproduce important features of the bulk phase diagram and therefore correctly predicts second phase formation in Al-Cu, and has been commonly used to investigate LC and grain boundary complexion nucleation in previous studies [8, 28]. However, this potential significantly underestimates the stacking fault energy of the alloy, leading to unrealistically larger partial dislocation spacing when compared to experimental observations. As such, a second angular dependent interatomic potential from Apostol and Mishin [29] that accurately predicts stacking fault energy and other important mechanical properties was used for deformation simulations. The Apostol and Mishin [29] interatomic potential was also tested in MC simulations; however, no platelet linear complexions were observed to form under the conditions studied, indicating its limitations in capturing such defect-stabilized structures.

To create the deformation simulation model, one of the platelet LC particles was isolated (i.e., other particles and defects were removed), followed by equilibration at 250 K using a micro-canonical ensemble (NVE) for 100 ps. Next, a new edge dislocation was introduced on the close-packed (111) plane away from the platelet precipitate by removing a half plane normal to the [110]

Burgers vector direction, followed by relaxation using a Nose–Hoover thermo/barostat at 250 K for 100 ps under zero pressure. Fig. 1 shows two examples where an edge dislocation has relaxed into a pair of Shockley partial dislocations. If the slip plane is located at the center of the platelet (Fig. 1(a)), classical precipitate-type interactions such as bowing or cutting will occur because the dislocations are physically blocked on their glide path. In contrast, if the slip plane is located just below the platelet (Fig. 1(b)), a LC-type reaction will occur where the compressive stress of the dislocations interacts with the platelet. The X-axis of the samples is oriented along the [110] direction (Burgers vector of the original dislocation), the Y-axis oriented along the $[1\bar{1}1]$ direction (slip plane normal), and the Z-axis oriented along the $[\bar{1}12]$ direction (line direction of the original dislocation). The simulation cells are approximately 34 nm long (X-direction), 25 nm tall (Y-direction), and 10 nm thick (Z-direction) and contain ~500,000 atoms. Non-periodic shrink-wrapped boundary condition are prescribed along the Y-direction and periodic boundary conditions along the X- and Z-directions.

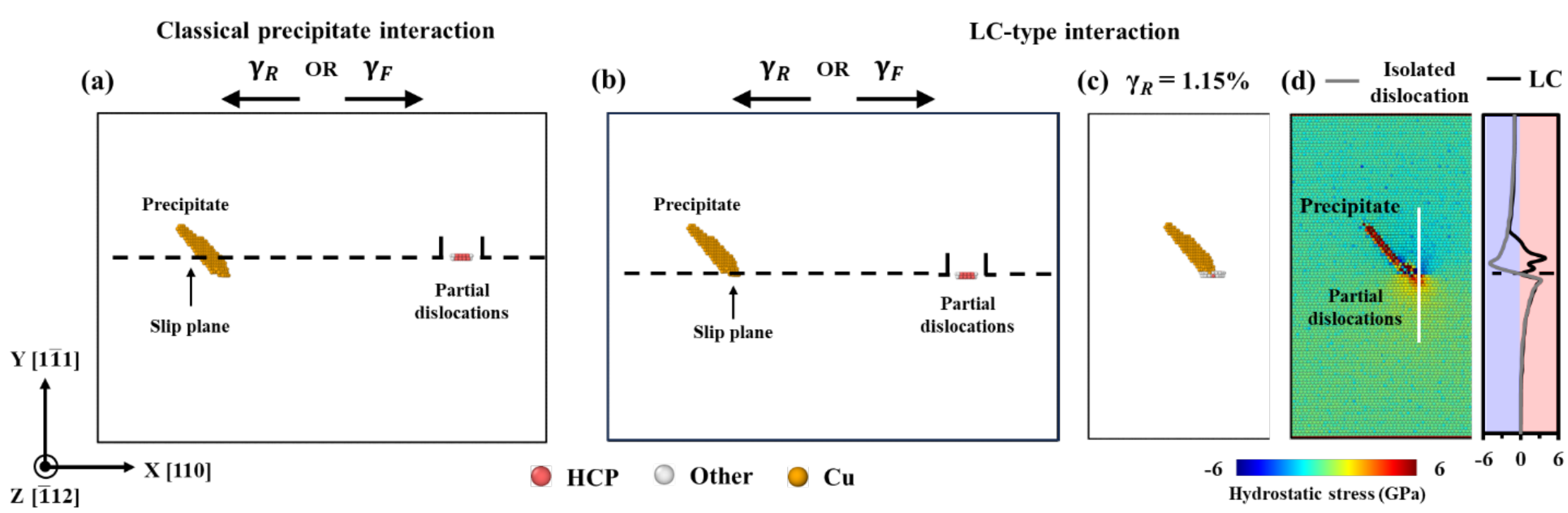

**Fig. 1. An equilibrated Al-Cu sample with a platelet precipitate and Shockley partial dislocation pair to simulate (a) classical precipitate and (b) LC-type interactions. Atoms are colored according to their local atomic structure (HCP, other, or Cu), while all of the Al atoms with FCC crystal structure have been removed for clarity. (c) An LC-type interaction with partial dislocations pinned at the precipitate during reverse shear**

**loading ($\gamma_R$), along with the (d) distribution of local hydrostatic stress along a vertical plane (denoted by a white line) in the sample, as compared with that of an isolated dislocation.**

To facilitate dislocation motion, progressive shear displacement under the canonical ensemble was applied to the Y-axis face in the X-direction to obtain a constant shear strain rate, with the two bottom and top layers of atoms (~0.5 nm) held fixed in the Y-direction to avoid rigid body rotation. The critical shear stress ($\tau_{yield}$) required for the dislocation to overcome the platelet precipitate during different types of interactions was measured and the applied strain rates ($\dot{\gamma}$) were varied from $5\times10^{6}$ s$^{-1}$ to $5\times10^{8}$ s$^{-1}$. The SRS parameter, *m*, is defined as [30, 31]:

$$m = \frac{\partial \, log(\tau_{yield})}{\partial \, log(\dot{\gamma})} \tag{1}$$

Since the precipitate is inclined at an angle of 30° with the dislocation line, the interaction between the precipitate and dislocation changes depending on the direction of dislocation motion. Thus, critical events were isolated for motion in each direction, being referred to as forward shear loading ($\gamma_F$) and reverse shear loading ($\gamma_R$). Fig. 1(c) shows an LC-type interaction where the partial dislocations are pinned below the precipitate during reverse shear loading ($\gamma_R$ = 1.15%) for $\dot{\gamma}$ = $5\times10^{8}$ s$^{-1}$, along with the distribution of local hydrostatic stress in the sample. As a result of the dislocation-platelet interactions, the compressive stresses of the dislocation are relaxed in Fig. 1(d), with the highly compressed region from the original dislocation (grey line) being reduced and replaced by a tensile stress for the LC (black line), due to the volumetric and structural misfit of the Cu-rich platelet that redistributes the local dislocation stress field.

Fig. 2 shows both classical precipitate and LC-type interactions during forward and reverse loading conditions for $\dot{\gamma}$ = $5\times10^{8}$ s$^{-1}$, with the viewing direction looking down into the slip plane. During the classical precipitate interaction, the partial dislocations overcome the obstacle via either

Orowan looping (forward) or precipitate cutting (reverse) mechanism, as shown in Figs. 2(a) and (b), respectively. In each case, the local structure or environment of the obstacle is altered as the partial dislocations move past, leaving behind either a dislocation loop or cutting through the precipitate. In contrast, the precipitate retains its original structure during the LC-type interactions shown in Figs. 2(c) and (d). The fact that dislocation-precipitate interactions only occur due to long-range stress fields rather than direct interactions can explain this lack of damage. The interaction mechanisms shown in Fig. 2 were found to remain consistent for a given dislocation placement and deformation direction as shear strain rate was varied.

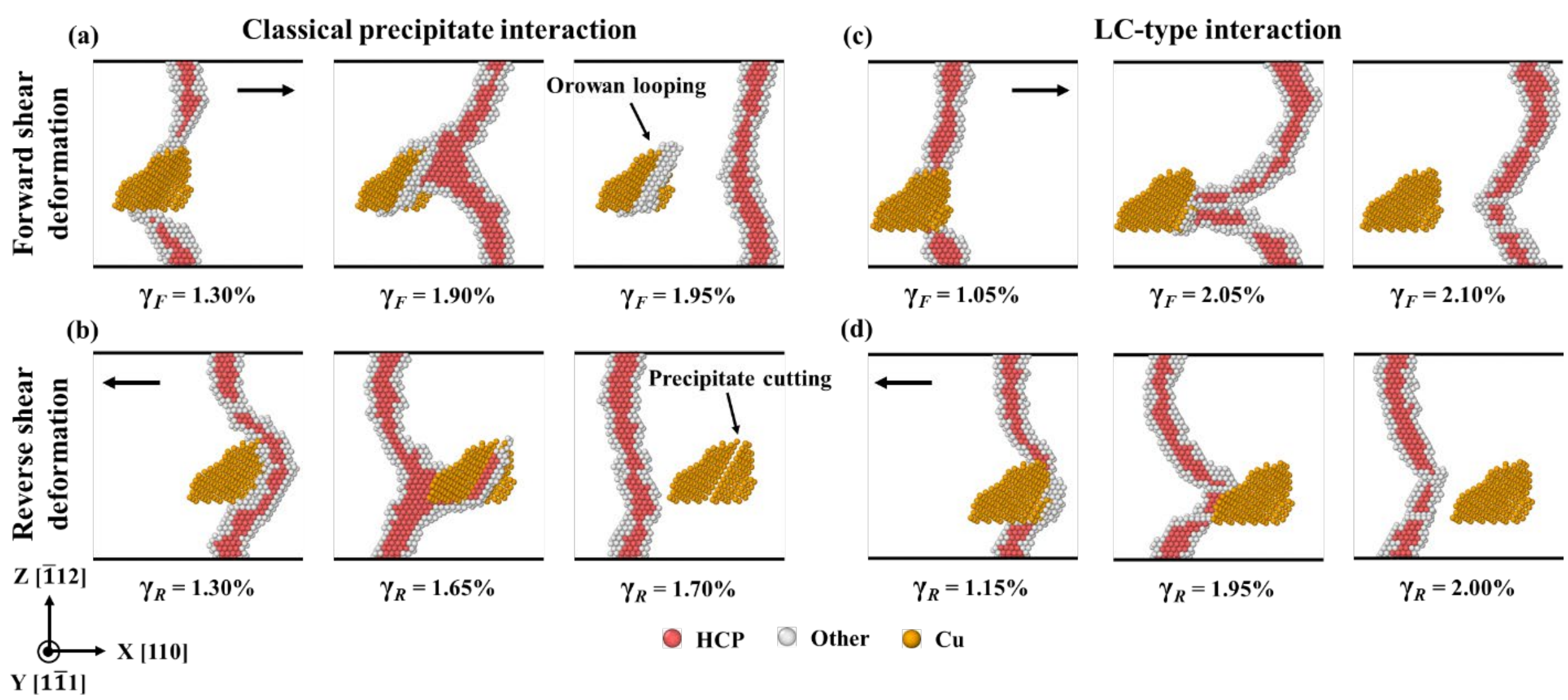

**Fig. 2. (a, b) Classical precipitate and (c, d) LC-type interactions at a representative shear strain rate of $5 \times 10^8$ $s^{-1}$, with simulation viewing direction along the slip plane normal. The partial dislocations overcome the obstacle via (a) Orowan looping or (b) precipitate cutting during the classical precipitate interactions. All of the FCC atoms have been removed for clarity.**

The mechanism by which dislocations overcome the precipitate during LC-type interactions is shown in detail in Figs. 3 and 4. In the classical case, the dislocation glides up to the precipitate, interacts with it, and then continues to move on the slip plane following well-known obstacle-interaction mechanisms. The classical precipitate interaction mechanisms and their dependence on parameters such as the shape, size, orientation of the precipitate, nature of the precipitate/matrix interface, and other variables have been examined in great detail in prior literature [32-36]. In contrast, the LC-type interactions involves a distinct sequence of events where the leading and trailing partials glide up to the platelet ($\gamma_F$ = 0.95%), with leading partial first climbing multiple atomic layers along the platelet ($\gamma_F$ = 1.00%), followed by the trailing partial climbing as well ($\gamma_F$ = 1.05%). These climbed segments then interact and recombine into a full dislocation, with recombination initiating on the side of the platelet closer to the dislocation line ($\gamma_F$ = 1.40%) and subsequently on the opposite side due to the platelet's 30° inclination with the dislocation line. Maximum climb occurs at $\gamma_F$ = 1.60%, where full dislocation character is observed across the platelet. As the sample is deformed further, the full dislocation dissociates back into Shockley partials ($\gamma_F$ = 1.90-2.00%) which subsequently climb downward one atomic layer at a time ($\gamma_F$ = 2.05%), while the dislocation segments away from the precipitate bow away from the obstacle. Eventually, the shifted dislocation segment returns to its original slip plane and resume glide as it breaks away from the obstacle.

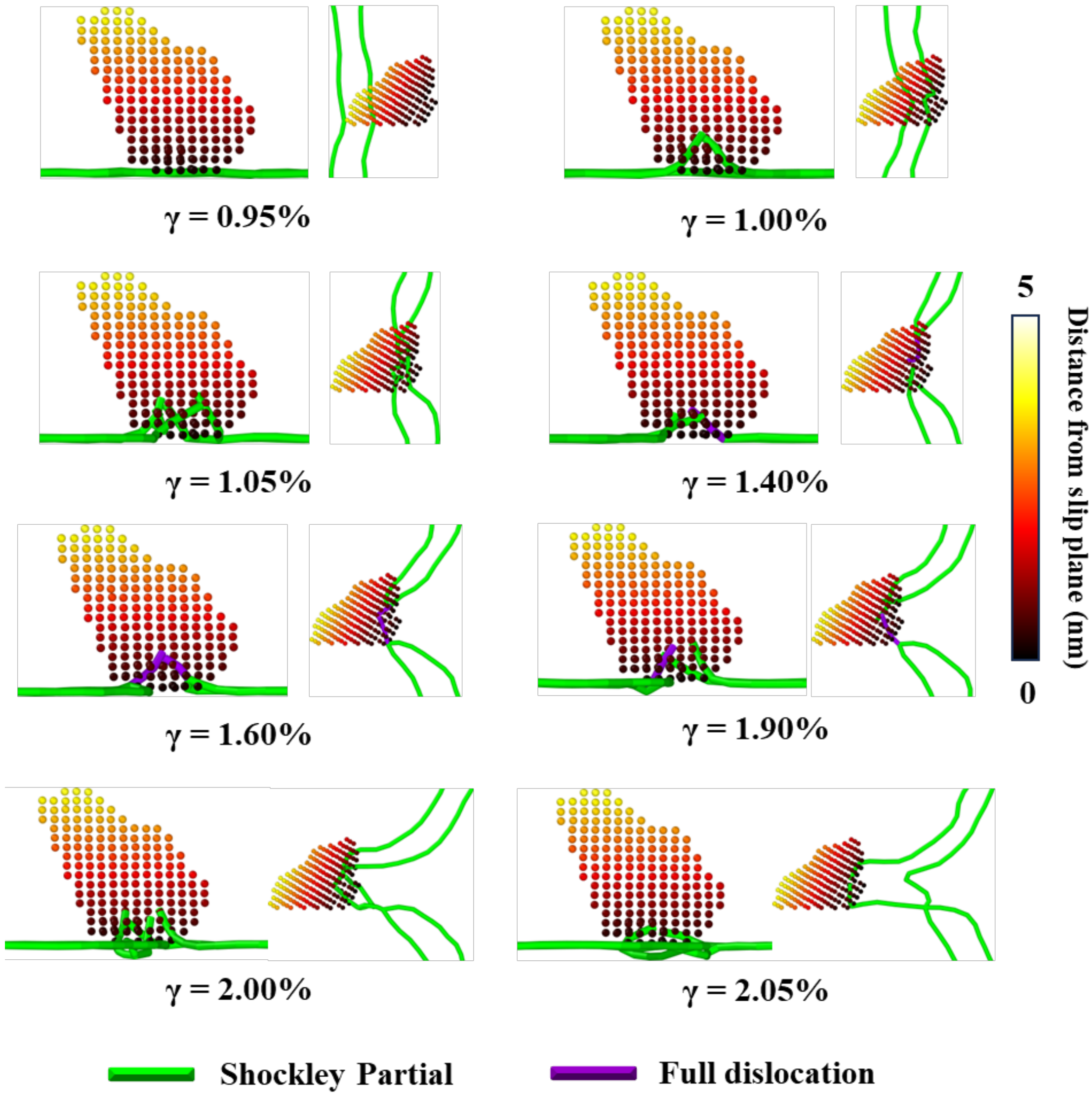

**Fig. 3. The dislocation climb process during LC-type interaction process at different shear strains for a forward shear strain rate of $5\times10^8$ $s^{-1}$, with simulation viewing directions facing the platelet and from the top. All of the FCC atoms have been removed for clarity. The platelet precipitate atoms are colored according to their distance above the slip plane.**

Fig. 4 shows the dislocation in more detail at its point of maximum climb during an LC-type interaction ($\gamma_F$ = 1.60%) and their dependence on precipitate morphology. Edge-on and top views of the precipitate are, with the atoms in the platelet precipitate colored according to their distance above the slip plane. The dislocation segments interacting with the precipitate recombine to form a full dislocation and shift out of their glide plane to move multiple atomic layers along the precipitate-matrix interface, whereas the dislocation segments away from the precipitate remain as a partial dislocation pair in the original slip plane. Analysis of the dislocation character with DXA shows that the shifted dislocation segments primarily have edge character (~80%) yet a minority of screw character (~20%) remains, with the primary edge character needed to facilitate the climb mechanism. This recombination and climb upwards into the obstacle, resulting in a 3D dislocation morphology, is unique for LC-type interactions. While climb can occur near traditional precipitates, the motion occurs through bulk regions to avoid the obstacle [37, 38], rather than along the obstacle interface as shown here.

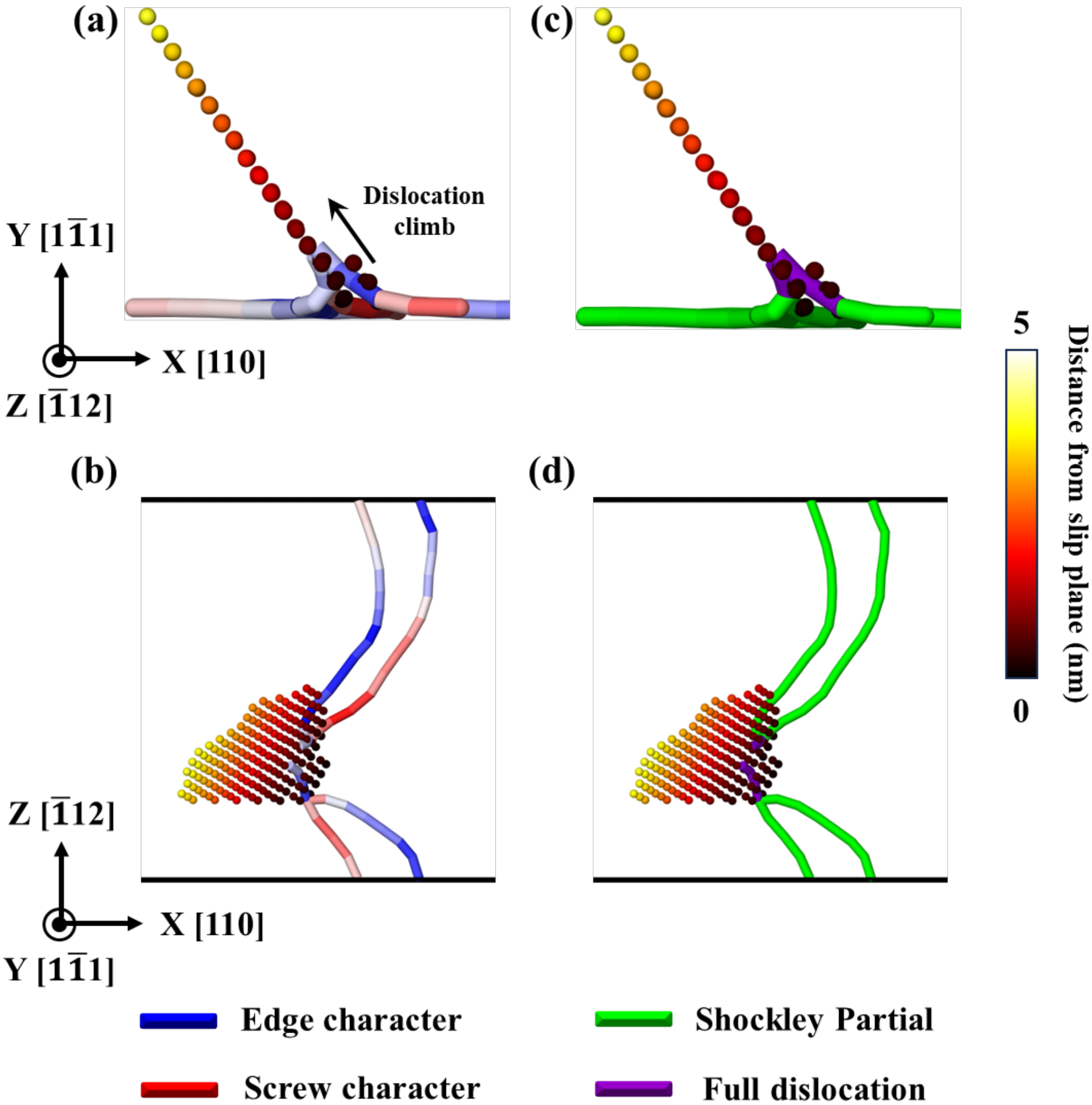

**Fig. 4. Analysis of a dislocation that has climbed into an LC, with the dislocation line colored according to the (a, b) dislocation character and (c, d) dislocation type. All of the FCC Al atoms are removed for clarity and Cu atoms are colored according to their distance from the slip plane. The edge-on views of the precipitate show that the partial dislocation segments recombine to form a full dislocation with primarily edge character and then climb along the precipitate-matrix interface out of the slip plane. The top view along the slip plane normal shows that the dislocation segments away from the precipitate remain as Shockley partial dislocations.**

Dislocation climb is an extremely rate-sensitive mechanism [39], suggesting that strengthening from LC-type interactions will depend strongly on deformation rate. The stress-strain responses for classical precipitate and LC-type interactions at three different reverse shear strain rates are shown in Fig. 5(a) and (b), respectively. The critical shear stress (denoted by black arrows) for LC-type interaction increases much more rapidly with increasing strain rate, reflecting the role of dislocation climb mechanism. We note that each curve contains an earlier feature that is associated with the dislocation moving forward to become stuck at the obstacle, but this is not the event of interest. Fig. 5(c) and (d) compiles all of the computed values for $\tau_{yield}$ at different applied shear strain rates for both classical precipitate and LC-type interactions. For classical precipitate interactions, yield strength is a relatively weak function of strain rate in Fig. 5(c). An $m$ value of 0.02 is measured for both forward ($\gamma_F$) and reverse ($\gamma_R$) deformation, giving a value that is consistent with previous reports for traditional Al alloys [40]. During LC-type interactions, $\tau_{yield}$ increases at a much higher rate with increasing applied shear strain rate in Fig. 5(d). Consequently, the measured values of $m$ are 0.08 and 0.11, or 4-5.5 times higher than the classical precipitate interactions. The only other literature data for Al that reports such a high SRS value is for thin film samples with nanocrystalline grain structures, with the higher $m$ values attributed to grain boundary processes dominating plasticity [41]. It is important to note that the $m$ values reported here are computed under constant strain rate loading conditions in MD, and thus represent an effective rate dependence rather than a direct equivalent to experimental $m$ values, which are typically obtained under constant-stress conditions. A more rigorous treatment of activation under ramped loading can be found in Zhu et al. [42, 43]. Additionally, the strain rate sensitivities reported here correspond to a single dislocation-obstacle interaction and therefore should not be expected to match experimental values that emerge from an ensemble of interacting obstacles [44-46]. Instead, these results highlight the relative enhancement in rate sensitivity due to linear complexions compared to conventional precipitate interactions.

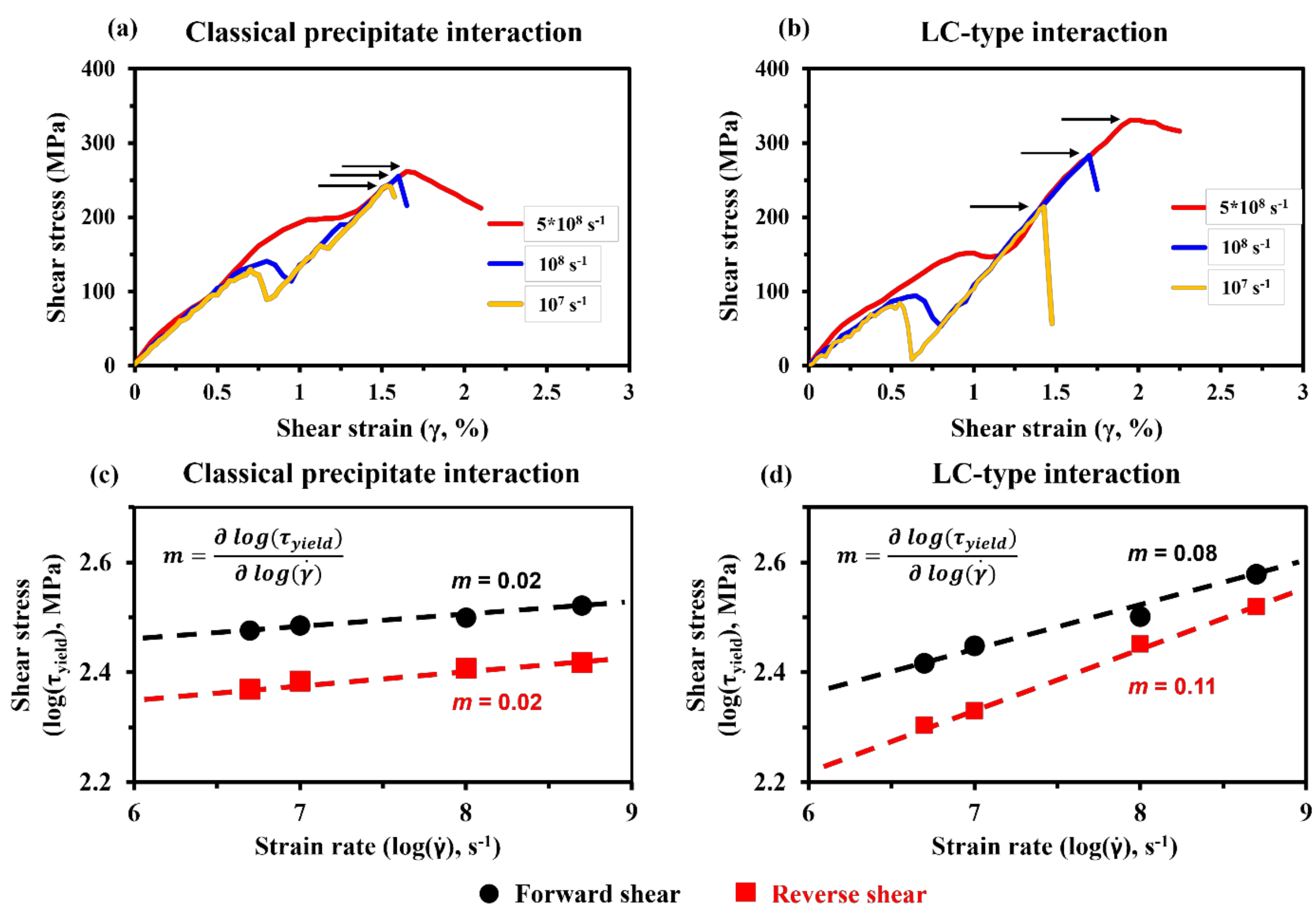

**Fig. 5. Stress-strain plots for (a) classical precipitate and (b) LC-type interaction process at three different reverse shear strain rates. The critical plastic flow event is denoted by black arrows for each curve. The critical shear stress ($\tau_{yield}$) required during (a) classical precipitate and (b) LC-type interactions for dislocation breakaway from the precipitate, presented as a function of the applied shear strain rates. The measured SRS is 4-5.5 times higher for the LC-type interactions.**

From a broader perspective, the discovery of platelet-shaped LCs in Al-Cu alloys introduces a novel microstructural feature that fundamentally alters the deformation behavior and mechanical response of these materials. Unlike classical GP zones, which form independently of dislocations [10, 11], LCs nucleate and stabilize directly in response to the local stress fields of dislocations. These stresses drive the evolution of the LCs into faceted platelet configurations that are energetically favorable and beneficial for mechanical response. These platelet LCs not only restrict dislocation glide but do so via mechanisms that deviate from traditional bowing or cutting models, instead requiring dislocation climb for further plastic deformation. These LCs allow dislocation climb to become important even at low temperatures, while it had previously only been observed to be critical under high-temperature creep conditions [22]. As a result, Al-Cu alloys containing platelet LCs should exhibit significantly higher SRS than those of conventional precipitate-strengthened Al alloys. This is particularly noteworthy because Al-Cu alloys typically show very low or negligible SRS at room temperature, with *m* values near zero due to the ease of dislocation glide and low Peierls barriers [13, 14]. Furthermore, an enhanced SRS implies improved resistance to dynamic loading, becoming much stronger when rapidly loaded. Even at quasi-static conditions, an enhancement in *m* as substantial as reported here would also imply the possibility to achieve large uniform tensile elongations, since high rate sensitivity can prolong the plastic strain regime prior to develop a necking instability, as described by the Hart criterion [47, 48]. This may be particularly attractive for Al alloys, which classically have a limited strain hardening capacity and low values of *m*.

In summary, the effect of different strain rates on plasticity associated with platelet array LCs was examined using MD simulations. For platelet LCs, the dislocation segments interacting with the precipitate recombine from a partial dislocation pair into a full dislocation near the platelet and then shift from the original slip plane, climbing a few atomic layers along the precipitate. Plasticity therefore requires climb in the opposite direction before the dislocation can break away from the precipitate and move freely. The simulated Al-Cu alloys with platelet LCs exhibited SRS values that are approximately five times higher than traditional precipitation strengthening mechanisms. As a whole, platelet array LCs represent a class of defect states that hold great promise for improving the impact strength of engineering alloys and provide a new pathway for defect engineering in light metal alloys.

## Availability of data and materials

The data that support the findings of this study are available within the article.

## Competing interests

The authors declare that they have no known competing financial interests or personal relationships that could have appeared to influence the work reported in this paper.

## Funding

This research was sponsored by the Army Research Office under Grant Number W911NF-21-1-0288. The views and conclusions contained in this document are those of the authors and should not be interpreted as representing the official policies, either expressed or implied, of the Army Research Office or the U.S. Government. The U.S. Government is authorized to reproduce and distribute reprints for Government purposes notwithstanding any copyright notation herein.

## Authors' contributions

**Pulkit Garg**: Conceptualization, Methodology, Investigation, Writing – original draft, Writing – review & editing. **Daniel S. Gianola**: Conceptualization, Writing – review & editing,

Funding acquisition. **Timothy J. Rupert**: Conceptualization, Writing – review & editing, Supervision, Project administration, Funding acquisition.

**Acknowledgments**

Not Applicable.

**References**

[1] M. Kuzmina, M. Herbig, D. Ponge, S. Sandlöbes, D. Raabe, Linear complexions: Confined chemical and structural states at dislocations, Science 349(6252) (2015) 1080-1083.
[2] A. Kwiatkowski da Silva, D. Ponge, Z. Peng, G. Inden, Y. Lu, A. Breen, B. Gault, D. Raabe, Phase nucleation through confined spinodal fluctuations at crystal defects evidenced in Fe-Mn alloys, Nature Communications 9(1) (2018) 1137.
[3] A. Kwiatkowski da Silva, G. Leyson, M. Kuzmina, D. Ponge, M. Herbig, S. Sandlöbes, B. Gault, J. Neugebauer, D. Raabe, Confined chemical and structural states at dislocations in Fe-9wt%Mn steels: A correlative TEM-atom probe study combined with multiscale modelling, Acta Materialia 124 (2017) 305-315.
[4] V. Turlo, T.J. Rupert, Linear Complexions: Metastable Phase Formation and Coexistence at Dislocations, Physical Review Letters 122(12) (2019) 126102.
[5] V. Turlo, T.J. Rupert, Interdependent Linear Complexion Structure and Dislocation Mechanics in Fe-Ni, Crystals 10(12) (2020) 1128.
[6] V. Turlo, T.J. Rupert, Dislocation-assisted linear complexion formation driven by segregation, Scripta Materialia 154 (2018) 25-29.
[7] G.R. Odette, N. Almirall, P.B. Wells, T. Yamamoto, Precipitation in reactor pressure vessel steels under ion and neutron irradiation: On the role of segregated network dislocations, Acta Materialia 212 (2021) 116922.
[8] V. Turlo, T.J. Rupert, Prediction of a wide variety of linear complexions in face centered cubic alloys, Acta Materialia 185 (2020) 129-141.
[9] H.C. Howard, W.S. Cunningham, A. Genc, B.E. Rhodes, B. Merle, T.J. Rupert, D.S. Gianola, Chemically ordered dislocation defect phases as a new strengthening pathway in Ni–Al alloys, Acta Materialia 289 (2025) 120887.
[10] C. Liu, S.K. Malladi, Q. Xu, J. Chen, F.D. Tichelaar, X. Zhuge, H.W. Zandbergen, In-situ STEM imaging of growth and phase change of individual CuAlX precipitates in Al alloy, Scientific reports 7(1) (2017) 2184.
[11] S.P. Ringer, K. Hono, Microstructural Evolution and Age Hardening in Aluminium Alloys: Atom Probe Field-Ion Microscopy and Transmission Electron Microscopy Studies, Materials Characterization 44(1) (2000) 101-131.
[12] P. Garg, D.S. Gianola, T.J. Rupert, Strengthening from dislocation restructuring and local climb at platelet linear complexions in Al-Cu alloys, Journal of Materials Science: Materials Theory 8(1) (2024) 9.

[13] M.S. Mohebbi, A. Akbarzadeh, Development of equations for strain rate sensitivity of UFG aluminum as a function of strain rate, International Journal of Plasticity 90 (2017) 167-176.
[14] S.L. Yan, H. Yang, H.W. Li, X. Yao, Variation of strain rate sensitivity of an aluminum alloy in a wide strain rate range: Mechanism analysis and modeling, Journal of Alloys and Compounds 688 (2016) 776-786.
[15] A.S. Khan, H. Liu, Variable strain rate sensitivity in an aluminum alloy: Response and constitutive modeling, International Journal of Plasticity 36 (2012) 1-14.
[16] R.C. Picu, G. Vincze, F. Ozturk, J.J. Gracio, F. Barlat, A.M. Maniatty, Strain rate sensitivity of the commercial aluminum alloy AA5182-O, Materials Science and Engineering: A 390(1) (2005) 334-343.
[17] E. Huskins, B. Cao, K. Ramesh, Strengthening mechanisms in an Al–Mg alloy, Materials Science and Engineering: A 527(6) (2010) 1292-1298.
[18] F. Kabirian, A.S. Khan, A. Pandey, Negative to positive strain rate sensitivity in 5xxx series aluminum alloys: Experiment and constitutive modeling, International Journal of Plasticity 55 (2014) 232-246.
[19] S. Yan, H. Yang, H. Li, G. Ren, Experimental study of macro–micro dynamic behaviors of 5A0X aluminum alloys in high velocity deformation, Materials Science and Engineering: A 598 (2014) 197-206.
[20] D.T. Hong Hue, V.-K. Tran, V.-L. Nguyen, L. Van Lich, V.-H. Dinh, T.-G. Nguyen, High strain-rate effect on microstructure evolution and plasticity of aluminum 5052 alloy nano-multilayer: A molecular dynamics study, Vacuum 201 (2022) 111104.
[21] H. Fan, Q. Wang, J.A. El-Awady, D. Raabe, M. Zaiser, Strain rate dependency of dislocation plasticity, Nature Communications 12(1) (2021) 1845.
[22] W.J. Kim, Y.K. Sa, H.K. Kim, U.S. Yoon, Plastic forming of the equal-channel angular pressing processed 6061 aluminum alloy, Materials Science and Engineering: A 487(1) (2008) 360-368.
[23] A.P. Thompson, H.M. Aktulga, R. Berger, D.S. Bolintineanu, W.M. Brown, P.S. Crozier, P.J. in 't Veld, A. Kohlmeyer, S.G. Moore, T.D. Nguyen, R. Shan, M.J. Stevens, J. Tranchida, C. Trott, S.J. Plimpton, LAMMPS - a flexible simulation tool for particle-based materials modeling at the atomic, meso, and continuum scales, Computer Physics Communications 271 (2022) 108171.
[24] D. Faken, H. Jónsson, Systematic analysis of local atomic structure combined with 3D computer graphics, Computational Materials Science 2(2) (1994) 279-286.
[25] A. Stukowski, V.V. Bulatov, A. Arsenlis, Automated identification and indexing of dislocations in crystal interfaces, Modelling and Simulation in Materials Science and Engineering 20(8) (2012) 085007.
[26] A. Stukowski, Visualization and analysis of atomistic simulation data with OVITO–the Open Visualization Tool, Modelling and Simulation in Materials Science and Engineering 18(1) (2009) 015012.
[27] Y. Cheng, E. Ma, H. Sheng, Atomic level structure in multicomponent bulk metallic glass, Physical review letters 102(24) (2009) 245501.
[28] Y. Hu, T.J. Rupert, Atomistic modeling of interfacial segregation and structural transitions in ternary alloys, Journal of Materials Science 54(5) (2019) 3975-3993.
[29] F. Apostol, Y. Mishin, Interatomic potential for the Al-Cu system, Physical Review B 83(5) (2011) 054116.

[30] N. Tahreen, D.L. Chen, M. Nouri, D.Y. Li, Effects of aluminum content and strain rate on strain hardening behavior of cast magnesium alloys during compression, Materials Science and Engineering: A 594 (2014) 235-245.
[31] W. Wang, G. Wang, Y. Hu, G. Guo, T. Zhou, Y. Rong, Temperature-dependent constitutive behavior with consideration of microstructure evolution for as-quenched Al-Cu-Mn alloy, Materials Science and Engineering: A 678 (2016) 85-92.
[32] E. Orowan, Symposium on internal stresses in metals and alloys, Institute of Metals, London 451 (1948).
[33] A.J. Ardell, Precipitation hardening, Metallurgical Transactions A 16(12) (1985) 2131-2165.
[34] V. Gerold, On the structures of Guinier-Preston zones in AlCu alloys introductory paper, Scripta Metallurgica 22(7) (1988) 927-932.
[35] C.V. Singh, D.H. Warner, Mechanisms of Guinier–Preston zone hardening in the athermal limit, Acta Materialia 58(17) (2010) 5797-5805.
[36] I. Adlakha, P. Garg, K.N. Solanki, Revealing the atomistic nature of dislocation-precipitate interactions in Al-Cu alloys, Journal of Alloys and Compounds 797 (2019) 325-333.
[37] Y. Xiang, D. Srolovitz, Dislocation climb effects on particle bypass mechanisms, Philosophical magazine 86(25-26) (2006) 3937-3957.
[38] C.V. Singh, A.J. Mateos, D.H. Warner, Atomistic simulations of dislocation–precipitate interactions emphasize importance of cross-slip, Scripta Materialia 64(5) (2011) 398-401.
[39] A. Vevecka-Priftaj, A. Böhner, J. May, H.W. Höppel, M. Göken, Strain rate sensitivity of ultrafine grained aluminium alloy AA6061, Materials Science Forum, Trans Tech Publ, 2008, pp. 741-747.
[40] S. Saroukhani, L.D. Nguyen, K.W.K. Leung, C.V. Singh, D.H. Warner, Harnessing atomistic simulations to predict the rate at which dislocations overcome obstacles, Journal of the Mechanics and Physics of Solids 90 (2016) 203-214.
[41] D.S. Gianola, D.H. Warner, J.F. Molinari, K.J. Hemker, Increased strain rate sensitivity due to stress-coupled grain growth in nanocrystalline Al, Scripta Materialia 55(7) (2006) 649-652.
[42] T. Zhu, J. Li, Ultra-strength materials, Progress in Materials Science 55(7) (2010) 710-757.
[43] T. Zhu, J. Li, A. Samanta, A. Leach, K. Gall, Temperature and strain-rate dependence of surface dislocation nucleation, Physical Review Letters 100(2) (2008) 025502.
[44] D. Zhang, R.C. Picu, Solute clustering in Al–Mg binary alloys, Modelling and Simulation in Materials Science and Engineering 12(1) (2003) 121.
[45] S. Xu, L. Xiong, Y. Chen, D.L. McDowell, Edge dislocations bowing out from a row of collinear obstacles in Al, Scripta Materialia 123 (2016) 135-139.
[46] S. Xu, L. Xiong, Y. Chen, D.L. McDowell, An analysis of key characteristics of the Frank-Read source process in FCC metals, Journal of the Mechanics and Physics of Solids 96 (2016) 460-476.
[47] E.W. Hart, Theory of the tensile test, Acta Metallurgica 15(2) (1967) 351-355.
[48] Y.M. Wang, E. Ma, Three strategies to achieve uniform tensile deformation in a nanostructured metal, Acta Materialia 52(6) (2004) 1699-1709.